# Noise from metallic surfaces – effects of charge diffusion

Carsten Henkel[1] and Baruch Horovitz[2]

[1]*Institut für Physik, Karl-Liebknecht-Str. 24-25,*
*Universität Potsdam, 14476 Potsdam, Germany and*

[2] *Department of Physics, Ben Gurion University, Beer Sheva 84105, Israel*

Non-local electrodynamic models are developed for describing metallic surfaces for a diffusive metal. The electric field noise at a distance $z_0$ from the surface is evaluated and compared with data from ion chips that show anomalous heating with a noise power decaying as $z_0^{-4}$. We find that high surface diffusion can account for the latter result.

## I. INTRODUCTION

A growing number of groups are using atom-chips or ion-chips to confine cold atoms, in anticipation of future applications and fundamental discoveries [1]. Progress in cold atom physics has led, however, to extreme sensitivity to electromagnetic noise from metallic surfaces since the chips contain inevitably metallic components. Atom chip experiments with conventional metal structures have demonstrated the existence of magnetic fluctuations way above the level of blackbody radiation [2, 3], as predicted by a theoretical study [4].

Experiments with ion chips have shown "anomalous heating", i.e. the ions absorb energy faster than expected from Nyquist noise of a good metal [5, 6, 7, 8]. The dependence on distance $z_0$ to the surface ($\sim z_0^{-4}$) also does not conform the expected form of Ref.4, proportional to $z_0^{-2}$. A tentative explanation are charge fluctuations in the surface of the metallic electrodes that enhance the electric field noise[5]. One of the motivations of this work is to develop a model for this process, based on the diffusive motion of charge carriers. Due to diffusion, the metallic medium responds in a nonlocal way to an applied field, which has been well studied in the past[9, 10]. The topic has recently attracted interest again to understand the Lifshitz-Casimir force between doped semiconductors [11, 12].

In the present work we develop a method to treat non-local electrodynamics (NLED) corresponding to a momentum-dependent dielectric function. In particular we allow for a coupling between longitudinal and transverse field fluctuations via the electrons' diffusion constant. Our approach is valid on length scales large compared to the electronic mean free path. Other small scales that are not resolved here are the Thomas-Fermi (or Debye-Hückel) screening length and the Fermi wavelength. Relevant scales are the distance of observation from a planar surface, the wavelength in the medium (skin depth) at a given frequency and the diffusive path length over one field period. We shall make the approximation that the vacuum wavelength is very large which is justified for the distances ($1-100\mu$m) and frequencies (100 kHz to 100 MHz) relevant for atom and ion chips.

We find that if the charge fluctuations are confined to the surface and have a high diffusion constant, then the anomalous heating observed in ion micro-traps can be accounted for. The reasoning for such a surface layer is either (i) the short charge screening length of the metal which is of the order of one atomic unit, or (ii) a distinct electron band existing at the surface. The latter scenario is known for a number of metals and has been recently confirmed via an observation of acoustic surface plasmons[13, 14].

The outline of the paper: in section II we review the method for evaluating electromagnetic noise and its relation to the ion heating data. In section III we present general properties of the NLED method for a normal metal allowing for dissipation and diffusion. In section IV we apply the method to electric noise above a diffusive surface layer, while in section V we evaluate the more conventional case of a continuous charge. The appendices complete the paper with a review of a local medium scattering problem and and a nonlocal calculation of magnetic noise.

## II. HEATING RATE AND NOISE

Consider a normal metal that occupies the half space at $z < 0$. The aim is to evaluate the fluctuations of the electric field $\mathbf{E}(\mathbf{r}, t)$ outside the metal, due to thermal or quantum fluctuations in the metal. The metal is in thermal equilibrium. An efficient procedure [4, 5, 15, 16] uses the following steps: (i) Introduce a source dipole $\mathbf{a}e^{-i\omega t}$ at $\mathbf{r}_0$ outside the metal and evaluate the wave emitted by the dipole, $\mathbf{E}^i(\mathbf{r}, t)$, which is the incident wave on the surface. (ii) Solve the scattering problem at the surface and find the reflected wave $\mathbf{E}^r(\mathbf{r}, t)$. This identifies the response function $\alpha_{i,j}$

$$E_i^r(\mathbf{r}_0, t) = \alpha_{i,j}(\omega) a_j e^{-i\omega t} \tag{1}$$

(iii) The final step is to use the Fluctuation Dissipation theorem (FDT) with the interaction $V_{int} = -\mathbf{E}(\mathbf{r}, t) \cdot \mathbf{a} e^{-i\omega t}$. The relevant type of FDT is related to the Golden rule that gives the transition rate $0 \to 1$ for a dipole

$$\begin{aligned} \Gamma_{0\to 1} &= \frac{1}{T\hbar^2} \sum_{f,i} \frac{e^{-E_i/k_B T}}{Z} \left| \int_0^T dt\, e^{i\omega_{10} t} \langle 1, f | H_{int} | i, 0 \rangle \right|^2 \\ &= \frac{a^2}{\hbar^2} \int d\tau\, e^{i\omega_{10}\tau} \langle E(t') E(t'+\tau) \rangle = \frac{a^2}{\hbar^2} \frac{2}{e^{\beta\omega_{10}} - 1} \operatorname{Im} \alpha_{i,j}(\omega_{10}) \end{aligned} \tag{2}$$

where we assume metal and field in thermal equilibrium at the inverse temperature $1/k_B T = \beta/\hbar$ (partition function $Z$) and sum over the final states $|f\rangle$. These states are actually polaritonic states including the excitations of the metallic medium (see, e.g., Ref.17).

Consider now the emitted wave from a unit charge oscillating at position $\mathbf{r}(t) = \mathbf{r}_0 + \mathbf{a} e^{-i\omega t}$. In the limit $\mathbf{a} \to 0$ the charge and current densities are

$$\begin{aligned} \rho(\mathbf{r}, t) &= -\mathbf{a} \cdot \boldsymbol{\nabla} \delta^3(\mathbf{r} - \mathbf{r}_0) e^{-i\omega t} \\ \mathbf{J}(\mathbf{r}, t) &= -i\omega \mathbf{a} \delta^3(\mathbf{r} - \mathbf{r}_0) e^{-i\omega t} \,. \end{aligned} \tag{3}$$

The emitted wave can be found via the vector potential $\mathbf{A}(\mathbf{r}, t) = \mathbf{A}(\mathbf{r}) e^{-i\omega t}$, which in

the Lorentz gauge is

$$\mathbf{A}(\mathbf{r}) = \int \mathbf{J}(\mathbf{r}') \frac{e^{i\omega|\mathbf{r}-\mathbf{r}'|/c}}{|\mathbf{r}-\mathbf{r}'|} d^3 r' = \frac{-i\omega}{c} \mathbf{a} \frac{e^{i\omega|\mathbf{r}-\mathbf{r}_0|/c}}{|\mathbf{r}-\mathbf{r}_0|}$$
$$= -\mathbf{a}\frac{i\omega}{c} 2\pi \int \frac{d^2 k}{(2\pi)^2} \frac{e^{i\mathbf{k}\cdot(\mathbf{r}-\mathbf{r}_0)-v_0|z-z_0|}}{v_0} \qquad (4)$$

where $\mathbf{r}_0 = (x_0, y_0, z_0)$, $\mathbf{k}$ is a 2-dimensional vector in $(x,y)$ directions and $v_0 = \sqrt{k^2 - (\omega/c)^2}$; for an outgoing wave we choose $\operatorname{Im} v_0 < 0$ and $\operatorname{Re} v_0 > 0$. The last line of (4) is known as the Weyl angular spectrum, and can be proven by showing that it solves $(\nabla^2 + \frac{\omega^2}{c^2})\mathbf{A}(\mathbf{r}) = -\frac{4\pi}{c}\mathbf{J}(\mathbf{r})$ with outgoing boundary condition.

The electric field in the vacuum layer $0 < z < z_0$ solves $(\nabla^2 + \frac{\omega^2}{c^2})\mathbf{E}(\mathbf{r}) = 0$, hence the incident wave has the form

$$\mathbf{E}^i(\mathbf{r}) = \int \frac{d^2 k}{(2\pi)^2} \mathbf{E}^i(\mathbf{k}) e^{i\mathbf{k}_i\cdot\mathbf{r}} \qquad (5)$$

where $\mathbf{k}_i = (\mathbf{k}, -iv_0)$. Since $\mathbf{E}^i(\mathbf{r}) = \frac{ic}{\omega}\boldsymbol{\nabla}\times\boldsymbol{\nabla}\times\mathbf{A}^i(\mathbf{r})$, we obtain from Eq.(4)

$$\mathbf{E}^i(\mathbf{k}) = -2\pi\mathbf{k}_i\times\mathbf{k}_i\times\mathbf{a}\frac{e^{-i\mathbf{k}_i\cdot\mathbf{r}_0}}{v_0} = -2\pi[(\mathbf{k}_i\cdot\mathbf{a})\mathbf{k}_i - \frac{\omega^2}{c^2}\mathbf{a}]\frac{e^{-i\mathbf{k}_i\cdot\mathbf{r}_0}}{v_0}. \qquad (6)$$

We are left then with the task of solving a scattering problem and identifying the response $\alpha_{i,j}(\omega)$, which via Eq.(2) will yield the heating rate.

## III. DIFFUSIVE METAL - GENERAL PROPERTIES

In this section we develop and apply a theory of nonlocal current-field response, equivalent to a $q$ dependent dielectric function. At low frequencies we use the constitutive $\mathbf{J}(\mathbf{E})$ relation as

$$\mathbf{J} = \sigma\mathbf{E} - D\boldsymbol{\nabla}\rho \qquad (7)$$

where $D$ is the diffusion constant and $\boldsymbol{\nabla}\cdot\mathbf{E} = 4\pi\rho$. Taking the gradient of (7) and using continuity $\boldsymbol{\nabla}\cdot\mathbf{J} = -\partial\rho/\partial t = i\omega\rho$ we obtain

$$\boldsymbol{\nabla}\cdot\mathbf{J} = 4\pi\sigma\rho - D\nabla^2\rho = i\omega\rho. \qquad (8)$$

In 3D this can be Fourier transformed to yield $4\pi\sigma + Dq^2 - i\omega = 0$, showing an overdamped plasma mode, as expected; it is also consistent with $\epsilon_L(q,\omega) = 0$ where the $q$ dependent longitudinal response is $\epsilon_L(q,\omega) = 1 + \frac{4\pi\sigma}{-i\omega+Dq^2}$.

Defining $\rho(\mathbf{r}) = \int\frac{d^2 k}{(2\pi)^2} e^{i\mathbf{k}\cdot\mathbf{r}}\rho(\mathbf{k},z)$, the solution of (8) has the form

$$\rho(\mathbf{k},z) = \rho(\mathbf{k},0)e^{v_1 z} \qquad v_1 = \sqrt{\frac{4\pi\sigma - i\omega}{D} + k^2}, \qquad \operatorname{Re} v_1 > 0. \qquad (9)$$

The typical scale of $v_1$ is determined by the Thomas-Fermi screening length $a_0 \equiv (D/4\pi\sigma)^{1/2}$ which in normal metals is $\approx 1\,\text{Å}$. Adopting a Drude model, we have $a_0^2 \approx \pi\hbar^2/(4me^2 k_F)$ with $k_F$ the Fermi wavenumber.

Eq.(7) is valid for small gradients, hence for a normal metal with a short $a_0$ we propose the following more general description. We assume that the charge has a surface component $\gamma(x,y)\delta(z)$ whose charge density $\gamma(x,y)$ will be determined self consistently. This surface layer takes care of the rapidly varying charge. In addition, we allow for a continuous charge $\rho(\mathbf{r})$ that extends into the bulk $z < 0$. The total charge is then

$$\rho_{tot}(\mathbf{r}) = \rho(\mathbf{r})\theta(-z) + \gamma(x,y)\delta(z) \tag{10}$$

The divergence $\boldsymbol{\nabla}\cdot\mathbf{E}$ involves a jump across the surface as well as a bulk term, hence

$$\boldsymbol{\nabla}\cdot\mathbf{E} = (E_z^{out} - E_z^{in})\delta(z) + (\boldsymbol{\nabla}\cdot\mathbf{E})\theta(-z) = 4\pi\rho(\mathbf{r})\theta(-z) + 4\pi\gamma(x,y)\delta(z) \tag{11}$$

and therefore

$$\begin{aligned} E_z^{out} - E_z^{in} &= 4\pi\gamma(x,y) \\ \boldsymbol{\nabla}\cdot\mathbf{E} &= 4\pi\rho(\mathbf{r}) \qquad z < 0 \end{aligned} \tag{12}$$

We generalize Eq.(7) to allow for surface diffusion

$$\mathbf{J} = (\sigma\mathbf{E} - D\boldsymbol{\nabla}\rho)\theta(-z) - D_s\boldsymbol{\nabla}_{||}\gamma(x,y)\delta(z) \tag{13}$$

where $\boldsymbol{\nabla}_{||}$ is a gradient with components parallel to the surface. We neglect here a conductive surface current as the surface layer is narrow. This current would be of the form $\sigma_s a_0 \mathbf{E}_{||}\delta(z)$ with $a_0$ the layer thickness and $\sigma_s$ of order $\sigma$. It would add to the following Eq.(14) a term $\sigma_s a_0 \boldsymbol{\nabla}_{||}\cdot\mathbf{E} \ll \sigma E_z$. For the latter estimate we consider the typical scale $k \sim 1/z_0$ and get a ratio $\approx a_0 k \sim a_0/z_0 \ll 1$. Eq.(13) is the simplest diffusion model that allows for enhanced surface diffusion relative to bulk (coefficient $D_s > D$) and that takes care of the broken isotropy at the surface.

Charge conservation yields

$$\boldsymbol{\nabla}\cdot\mathbf{J} = (\sigma\boldsymbol{\nabla}\cdot\mathbf{E} - D\nabla^2\rho)\theta(-z) - J_z^{in}\delta(z) - D_s\nabla^2_{||}\gamma(x,y)\delta(z) = i\omega[\rho\theta(-z) + \gamma(x,y)\delta(z)] \tag{14}$$

where $J_z^{in} = \sigma E_z^{in} - D\partial_z\rho(0)$ is the bulk current that flows into the charge layer. The $\theta(-z)$ terms in (14) reproduce the bulk form (8) while the $\delta(z)$ terms provide the following link between the bulk and surface charge densities:

$$J_z^{in} = \sigma E_z^{in} - D\partial_z\rho(0) = -D_s\nabla^2_{||}\gamma(x,y) - i\omega\gamma(x,y) \tag{15}$$

In the following we consider two limiting cases: (i) The charge layer model where in the bulk $\rho(\mathbf{r}) = 0$; the surface charge is then determined by (15), see Eq.(21) below. (ii) The second case is the continuous charge model where $\gamma(x,y) = 0$. In the latter case Eq.(15) yields the obvious boundary condition $J_z^{in} = 0$.

We proceed to derive the boundary conditions that will be used in the two models. The wave equation for the electric field is (for model (i) replace $D$ by $D_s$ in the following). Maxwell's equations are

$$\begin{aligned} \boldsymbol{\nabla}\times\mathbf{E} &= \frac{i\omega}{c}B \\ \boldsymbol{\nabla}\times\mathbf{B} &= \frac{-i\omega}{c}E + \frac{4\pi}{c}\mathbf{J} \end{aligned} \tag{16}$$

Hence

$$\boldsymbol{\nabla} \times \boldsymbol{\nabla} \times \mathbf{E} = \frac{\omega^2}{c^2}(\mathbf{E} + \frac{4\pi i}{\omega}\mathbf{J}) . \tag{17}$$

We insert the current density (7) and eliminate $\rho_{\text{tot}}$ by $\boldsymbol{\nabla} \cdot \mathbf{E} = 4\pi\rho_{\text{tot}}$ that applies to all $z$ in Eq.(12), giving

$$\nabla^2 \mathbf{E} + \frac{\omega^2}{c^2}\epsilon(\omega)\mathbf{E} = (1 + \frac{iD\omega}{c^2})\boldsymbol{\nabla}(\boldsymbol{\nabla} \cdot \mathbf{E}) . \tag{18}$$

with $\epsilon(\omega) = 1 + 4\pi i\sigma/\omega$ the usual transverse dielectric function. To estimate the correction $D\omega/c^2$, we introduce the skin depth $\delta = c/\sqrt{2\pi\sigma\omega}$ as the length scale associated with Ohmic damping of the field inside the bulk $[1/\delta = (\omega/c)\text{Im}\sqrt{\epsilon(\omega)}]$. Usually, $D\omega/c^2 = 2(a_0/\delta)^2 \ll 1$ so that the main effect of the diffusive currents is via the boundary conditions.

We have already noticed the jump in $E_z$, Eq.(6). Due to the diffusive surface current, there is also a jump in the magnetic field, e.g.:

$$B_y^{in} - B_y^{out} = -\frac{4\pi D_s}{c}\partial_x \gamma \tag{19}$$

By applying the $x$-derivative and combining with the corresponding equation for $B_x$, we get

$$E_z^{out} - \epsilon E_z^{in} + \frac{4\pi D_s}{i\omega}\nabla_{\|}^2 \gamma(x,y) - \frac{4\pi D}{i\omega}\partial_z \rho(0) = 0 . \tag{20}$$

which can also be derived by combining Eqs. (12,15). Finally, with the usual argument that $\oint \mathbf{E} \cdot d\boldsymbol{\ell} = \int \boldsymbol{\nabla} \times \mathbf{E}\, d\mathbf{s} \to 0$ for a contour approaching the boundary, we deduce $E_{x,y}^{out} = E_{x,y}^{in}$. Other boundary conditions depend on the model for the charge layer.

## IV. CHARGE LAYER MODEL

The charge below the metal surface is represented here as a surface charge. The relevant material parameters are the surface diffusion constant $D_s$ and the bulk conductivity $\sigma$. The surface charge density $\gamma(x,y)$ is determined self-consistently from the boundary conditions. Setting the volume charge density $\rho(\mathbf{r}) = 0$ in Eq.(15), we find in $\mathbf{k}$-space

$$\gamma(\mathbf{k}) = \frac{i\sigma}{\omega + iD_s k^2}E_z^{in}(\mathbf{k}) = \frac{(\epsilon-1)/4\pi}{1 + iD_s k^2/\omega}E_z^{in}(\mathbf{k}) \tag{21}$$

Recall that the wave vector $\mathbf{k}$ is two-dimensional, with components parallel to the metal surface. Eq.(20) yields now

$$E_z^{out} = \tilde{\epsilon}E_z^{in}, \qquad \tilde{\epsilon} = \frac{\epsilon + iD_s k^2/\omega}{1 + iD_s k^2/\omega} . \tag{22}$$

The effect of diffusion is to introduce the correction $D_s k^2/\omega$, which for the important scale $k = 1/\delta$ defines the dimensionless parameter

$$D_0 \equiv \frac{D_s}{\omega\delta^2} = \frac{2\pi\sigma D_s}{c^2} = \frac{8\pi^2 a_0^2 \lambda^2}{\delta^4}\frac{D_s}{D} . \tag{23}$$

6Taking typical values of room temperature metal conductivities and diffusion constants in the bulk, and assuming $D_s \approx D$, we find $D_0 \approx 1$. We note, however, that the surface diffusion $D_s$ is not well known. Furthermore, at lower temperatures $D_0$ is significantly enhanced.

To complete the boundary conditions, we recall that $\nabla \cdot \mathbf{E} = i\mathbf{k} \cdot \mathbf{E}_{\parallel} + \partial_z E_z$ where the tangential vector $\mathbf{E}_{\parallel}$ is continuous across the interface. Considering that $\nabla \cdot \mathbf{E} = 0$ at $z \neq 0$, we thus find

$$\partial_z E_z^{out} - \partial_z E_z^{in} = 0. \tag{24}$$

Integrating the $x$ component of (18) over the surface layer leads to

$$\partial_z E_x^{out} - \partial_z E_x^{in} = (1 + \frac{iD\omega}{c^2})ik_x\gamma = (1 + \frac{iD_s\omega}{c^2})\frac{\epsilon - 1}{1 + \frac{iD_s k^2}{\omega}}ik_x E_z^{in}, \tag{25}$$

which can also be found from the jump in $B_y$ due to the diffusive surface current, Eq.(19).

We proceed now to solve the scattering problem. At $z \neq 0$, $\mathbf{E}$ solves

$$(\nabla^2 + \frac{\omega^2}{c^2})\mathbf{E}(\mathbf{r}) = 0$$
$$(\nabla^2 + \frac{\omega^2}{c^2}\epsilon(\omega))\mathbf{E}(\mathbf{r}) = 0 \tag{26}$$

The general form of the incident, reflected and transmitted waves is then

$$\mathbf{E}(\mathbf{r}) = \int \frac{d^2k}{(2\pi)^2}[\mathbf{E}^i(\mathbf{k})e^{i\mathbf{k}\cdot\mathbf{r}+v_0 z} + \mathbf{E}^r(\mathbf{k})e^{i\mathbf{k}\cdot\mathbf{r}-v_0 z}] \quad z > 0$$
$$\mathbf{E}(\mathbf{r}) = \int \frac{d^2k}{(2\pi)^2}\mathbf{E}^t(\mathbf{k})e^{i\mathbf{k}\cdot\mathbf{r}+vz} \quad z < 0 \tag{27}$$

where $v = \sqrt{k^2 - \frac{\omega^2}{c^2}\epsilon(\omega)}$ and the sign is chosen so that transmission decays, i.e. $\mathrm{Re}(v) > 0$. Boundary conditions for the $z$ component yield

$$\tilde{\epsilon} E_z^t = E_z^i + E_z^r$$
$$v E_z^t = v_0 (E_z^i - E_z^r) \tag{28}$$

so that

$$E_z^r(\mathbf{k}) = \frac{\tilde{\epsilon} v_0 - v}{\tilde{\epsilon} v_0 + v} E_z^i(\mathbf{k}). \tag{29}$$

We observe that this has the same form as the Fresnel reflection coefficient in p-polarization (transverse magnetic), but with a nonlocal, effective permittivity $\tilde{\epsilon} = \tilde{\epsilon}(k,\omega)$ in place of the local value. The imaginary part of this nonlocal reflection coefficient is plotted in Fig.1 (thick solid line). This quantity determines the spectral noise power of electric fields polarized normal to the surface (see Section II). A clear enhancement at intermediate $k$-vectors is seen.

Working out the angular integration in Eq.(27), the reflected field, evaluated at the source position, becomes

$$E_z^r(\mathbf{r}_0) = a_z \int_0^\infty dk\, k^3 \frac{\tilde{\epsilon} v_0 - v}{\tilde{\epsilon} v_0 + v} \frac{e^{-2v_0 z_0}}{v_0}, \tag{30}$$

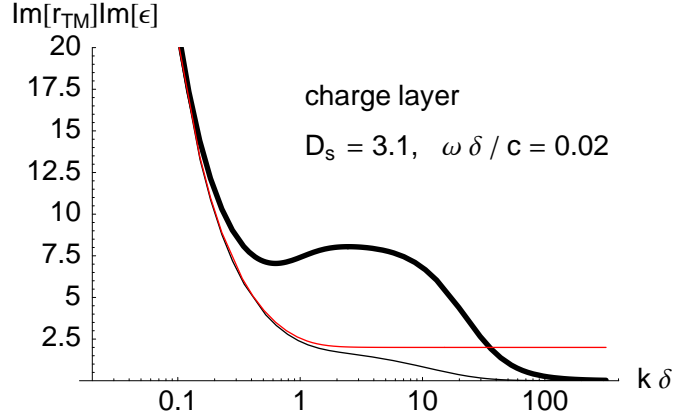

FIG. 1: Imaginary part of TM-polarized reflection coefficient vs. wave vector $k$ for a charge layer with lateral diffusion. Thick line: charge layer with lateral diffusion constant $D_s$, the reflection coefficient is defined as the ratio $E_z^r(\mathbf{k})/E_z^i(\mathbf{k})$ in Eq.(29). Thin black line: continuous charge model, Eq.(49), with bulk diffusion constant $D = D_s$. Thin red line (mainly constant): local calculation, Eq.(A6). $k$ is scaled to the inverse skin depth $1/\delta$. The imaginary part of the reflection coefficient is multiplied by $\mathrm{Im}\,\epsilon$ because in the local limit, $\mathrm{Im}\,r_p \to 2/\mathrm{Im}\,\epsilon$ at large $k$. The parameter choice $\omega\delta/c = 0.02$ is arbitrary.

which is the generalization of previous results obtained in the local approximation (see Appendix A). The surface response function $\alpha_{zz}(\omega)$ can be read off here, according to Eq.(1).

We now compute the behavior of $\mathrm{Im}\,\alpha_{zz}(\omega)$ at short distances $z_0 \ll \lambda$ which determines the spectral strength of electric near fields. Note that the typical $k$-vectors in the integral (30) are of the order of $k \sim 1/z_0$ where $v_0 \approx k$ because $z_0$ is much smaller than the wavelength. In the regime $\delta \ll z_0 \ll \lambda$ we have $v \approx (1-i)/\delta$. We also neglect the $D_s$ term in $\epsilon - iD_s k^2/\omega = 1 + \frac{4\pi i\sigma}{\omega}(1 - \frac{D_s k^2}{4\pi\sigma})$ since $D_s k^2/4\pi\sigma \sim a_0^2/z_0^2 \ll 1$. Hence

$$\frac{\tilde{\epsilon}v_0 - v}{\tilde{\epsilon}v_0 + v} \approx 1 - \frac{2}{k} \cdot \frac{1 + iD_s k^2/\omega}{4\pi i\sigma/\omega} \cdot \frac{1-i}{\delta} \qquad (31)$$

where the last terms are small, $\frac{\omega}{k\sigma\delta} \approx \frac{z_0\delta}{8\pi^2\lambda^2} \ll 1$, $\frac{D_s k}{4\pi\sigma\delta} \approx \frac{a_0^2}{z_0\delta} \ll 1$ corresponding to $a_0^2 \ll z_0\delta \ll \lambda^2$. The imaginary part has then two terms

$$\mathrm{Im}\,\alpha_{zz} = \frac{\omega}{8\pi\sigma}\frac{1}{z_0^2\delta}\left(1 + D_0\frac{3\delta^2}{2z_0^2}\right) \qquad (32)$$

where we used the dimensionless $D_0$ defined in Eq.(23). Note the distinct $z_0^4$ power when the diffusion term dominates.

In the regime $z_0 \ll \delta$ we use $v \approx k(1 - i/\delta^2 k^2)$ so that

$$\frac{\tilde{\epsilon}v_0 - v}{\tilde{\epsilon}v_0 + v} \approx 1 - \frac{2}{\tilde{\epsilon}} + \frac{2i}{\delta^2 k^2 \tilde{\epsilon}} \qquad (33)$$

Similar calculations give

$$\mathrm{Im}\,\alpha_{zz} = \frac{\omega}{8\pi\sigma}\frac{1}{z_0^3}(1 + D_0) \qquad (34)$$

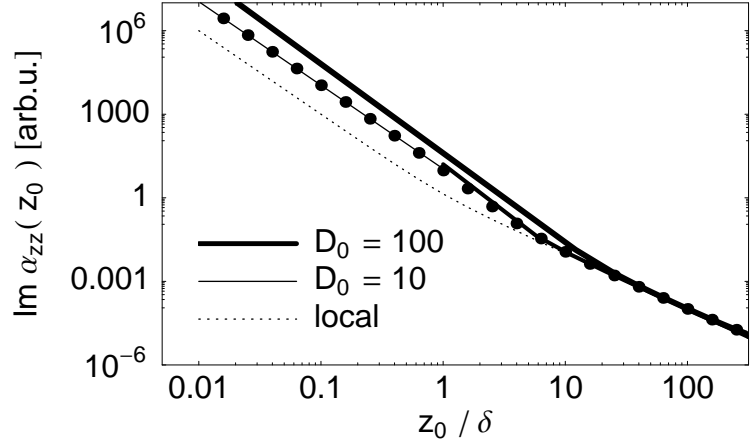

FIG. 2: Electric field noise for the surface charge model, vs. observation distance $z_0$ (scaled to the skin depth). We plot the quantity $\operatorname{Im}\alpha_{zz}(z_0,\omega)$ at fixed frequency and arbitrarily scaled. The surface diffusion constant is proportional to the dimensionless parameter $D_0$. Dotted line: local calculation by numerical integration of Eq.(A8); solid lines and symbols: nonlocal calculations. For $D_0 = 10$, the symbols are found by numerically integrating the imaginary part of Eq.(30), the solid lines from the asymptotic formulas (32) (for $z > \delta$) and (34) (for $z < \delta$). $D_0 = 100$ (thick solid line): numerical integration. For this plot, we arbitrarily choose $\omega\delta/c = 2\pi\delta/\lambda = 10^{-6}$.

so that surface diffusion enhances the $1/z_0^3$ term. These formulas are illustrated in Fig.2 and compared to a numerical calculation based on the integral (30).

Consider next the scattering equations for the tangential field. Without loss of generality, we work with $E_x$ and have

$$E_x^i + E_x^r = E_x^t$$
$$v_0(E_x^i - E_x^r) - vE_x^t = \frac{\epsilon-1}{1+iD_sk^2/\omega}iK_xE_z^t = iK_x\frac{\epsilon-1}{1+iD_sk^2/\omega}\cdot\frac{2v_0}{\tilde{\epsilon}v_0+v}E_z^i \qquad (35)$$

where $D_s\omega/c^2 \ll 1$ is neglected. Inserting $E_x^i$ and $E_z^i$ from (6)

$$E_x^r(\mathbf{r}_0) = 2\pi\int\frac{d^2k}{(2\pi)^2}\frac{e^{-2v_0z_0}}{v_0}\left[\frac{v-v_0}{v+v_0}\tfrac{1}{2}(v_0^2-\frac{\omega^2}{c^2}) + \frac{k^2v_0^2(\epsilon-1)}{(1+iD_sk^2/\omega)(v+v_0)(\tilde{\epsilon}v_0+v)}\right] \qquad (36)$$

Using $\tilde{\epsilon}\approx\epsilon/(1+iD_sk^2/\omega)$ we have for the correction to the local result (A15)

$$\Delta E_x^r(\mathbf{r}_0) = \int_0^\infty dk\, k\frac{e^{-2v_0z_0}}{v_0}\frac{-ik^2v_0^2(\epsilon-1)vD_sk^2/\omega}{(v+v_0)(\epsilon v_0+v)[\epsilon v_0+v(1+iD_sk^2/\omega)]}. \qquad (37)$$

For $\delta \ll z_0 \ll \lambda$ we use $v\approx\frac{1-i}{\delta}\gg k$ and $\epsilon k \gg v, vD_sk^2/\omega$. For the imaginary part we need the next order term which comes from $v+k\approx v(1+\frac{k\delta}{1-i})$; this correction is larger than all other ones,

$$\frac{\delta}{z_0}\gg\frac{\delta z_0}{\lambda^2},\ \frac{8\pi^2a_0^2}{\delta z_0},\ \frac{1}{\epsilon}\approx\frac{\delta^2}{8\pi^2\lambda^2} \qquad (38)$$

so that

$$\text{Im}\Delta\alpha_{xx} = \text{Im}\int_0^\infty dk\, e^{-2kz_0}\frac{-iD_s k^4/\omega}{4\pi\sigma i/\omega}(1 - \frac{k\delta}{1-i}) = \frac{15}{16}\frac{a_0^2\delta}{z_0^6}. \qquad (39)$$

For $z_0 \ll \delta$ we use $v \approx k(1 - i/\delta^2 k^2)$ as the leading correction so that

$$\text{Im}\alpha_{xx} = \int_0^\infty dk\, e^{-2kz_0}\frac{-iD_s k^4/\omega}{4\pi i\sigma/\omega}(1 - \frac{i}{2\delta^2 k^2}) = \frac{1}{4}\frac{a_0^2}{\delta^2 z_0^3}. \qquad (40)$$

To summarize all the results, we have at large distance $\delta \ll z_0 \ll \lambda$

$$\begin{aligned}
\text{Im}\alpha_{zz} &= \frac{\omega}{8\pi\sigma}\frac{1}{z_0^2 \delta}(1 + \frac{3}{2}D_0\frac{\delta^2}{z_0^2}) \\
\text{Im}\alpha_{xx} &= \frac{\omega}{8\pi\sigma}\frac{1}{z_0^2 \delta}(1 + \frac{15}{16}D_0\frac{\delta^4}{z_0^4})
\end{aligned} \qquad (41)$$

while at short distance $z_0 \ll \delta$ we have

$$\begin{aligned}
\text{Im}\alpha_{zz} &= \frac{\omega}{8\pi\sigma}\frac{1}{z_0^3}(1 + D_0) \\
\text{Im}\alpha_{xx} &= \frac{\omega}{16\pi\sigma}\frac{1}{z_0^3}(1 + D_0).
\end{aligned} \qquad (42)$$

The diffusion factor $D_0$ is seen to increase the fluctuations, in particular it introduces novel power laws that apply in intermediate distance ranges.

This result may account for the anomalous heating observed in cold ion experiments [5, 6, 7, 8]. In fact, the surface diffusion coefficient determines a length scale $\ell_D \sim (D_s/\omega)^{1/2}$ for the surface charge. Our charge layer model predicts an enhanced electric field noise for distances $z_0$ below $\delta$, with a $1/z_0^3$ dependence, an intermediate range $\delta \ll z_0 \ll \delta\sqrt{D_0} \sim \ell_D$ with a $1/z_0^4$ dependence, while at long distances $\ell_D \ll z_0 \ll \lambda$ the noise is insensitive to the charge layer. Both the enhanced noise and the power law $1/z_0^4$ qualitatively agree with the observed ion heating, provided one takes a large value $D_0 \gg 1$ for the normalized surface diffusion.

We note that regarding one point, these estimates are at variance with the previously invoked picture of patch charge fluctuations[5]: the typical patch size has been thought to be much smaller than $z_0$ to account for the $1/z_0^4$ law. Here, the increase in noise sets in if the diffusive length $\ell_D$ is larger than $z_0$. It would be interesting to investigate further what length actually best represents the size of a charge patch. Another candidate could be the (surface) mean free path $\ell_{\text{mfp}}$ with $D_s = \ell_{\text{mfp}}^2/\tau_s$ [$\tau_s$: surface scattering rate]. At the low frequencies relevant here, this scale is indeed much smaller, $\ell_{\text{mfp}} = \ell_D\sqrt{\omega\tau_s} \ll \ell_D$.

## V.  CONTINUOUS CHARGE MODEL

In this subsection we consider the solution for the charge density Eq.(9) as it is, i.e. without a surface charge. This is relevant if the electron density is very low so that $a_0$ is large compared with atomic scales, e.g. as in doped semiconductors [11].

The additional equation for $\rho(\mathbf{k}, z)$ necessitates an additional boundary condition: as discussed below Eq.(15), the $z$ component of the current vanishes at the surface,

$$J_z(\mathbf{k}, z = 0) = \sigma E_z^{in}(\mathbf{k}, 0) - D\partial_z \rho(\mathbf{k}, 0) = 0. \tag{43}$$

This determines the charge density, as discussed in Eq.(9) above:

$$\rho(\mathbf{k}, z) = \frac{\sigma}{Dv_1} E_z^{in}(\mathbf{k}, 0) e^{v_1 z}, \qquad z < 0 \tag{44}$$

and $\rho(\mathbf{k}, z) = 0$ for $z > 0$. There is a jump in the charge density, but no surface charge $\sim \delta(z)$.

Splitting the electric field in Eq.(18) into longitudinal and transverse parts, $\mathbf{E}_{L,T}$, we obtain

$$E_z^L(\mathbf{k}, z) = \frac{4\pi i D}{\omega \epsilon(\omega)} v_1 \rho(\mathbf{k}, 0) e^{v_1 z} \tag{45}$$

and

$$\mathbf{E}^T(\mathbf{k}, z) = \mathbf{E}^T(\mathbf{k}, 0) e^{vz} \tag{46}$$

As expected, these fields vary on two distinct length scales.

Since there is no surface charge, the electric field is continuous at $z = 0$:

$$E_z^{in} = E_z^{out} \tag{47}$$

but its derivative jumps

$$\partial_z E_z^{in} - \partial_z E_z^{out} = 4\pi \rho(\mathbf{k}, 0) \tag{48}$$

as can be also seen by integrating Eq.(18) across the surface. Note here that the diffusion term is confined to the inner part of the surface where $\rho(\mathbf{k}, z)$ is continuous; hence its contribution vanishes.

The solution of the scattering problem for the $z$ component is

$$E_z^r = \frac{\epsilon(\omega) v_0 - v - (\epsilon(\omega) - 1) k^2 / v_1}{\epsilon(\omega) v_0 + v + (\epsilon(\omega) - 1) k^2 / v_1} E_z^i \tag{49}$$

whose imaginary part is plotted in Fig.1 as thin solid line. It can be seen that diffusion in the bulk *reduces* the noise power on small scales (compared to the local calculation, red solid line), which is essentially a screening effect.

For $z_0 \ll \lambda$, we get a response function

$$\alpha_{zz} = \int_0^\infty dk\, k^3 \Big[ 1 - \frac{2v}{k\epsilon(\omega)} - (1 - \frac{1}{\epsilon(\omega)}) \frac{2k}{v_1} \Big] e^{-2kz_0} \tag{50}$$

where the first two terms in the bracket correspond to a local medium (see Appendix A). For both $\delta \ll z_0$ and $\delta \gg z_0$ we find that the corrections to the local form are of order $a_0/z_0$ and are therefore negligible. Significant changes only appear for $z \leq a_0$ where the divergent power laws are regularized, see, e.g. Ref.[18]. This regime is irrelevant, however, for atom and ion chips based on good conductors, because of the smallness of $a_0$.

For the $E_x$ response, we have the usual boundary condition $E_x^{in} = E_x^{out}$. Since there is no surface current, the magnetic field $B_y$ is continous as well, and we have $\partial_z E_x^{in} = \partial_z E_x^{out}$. Therefore

$$E_x^r = \frac{v_0 - v}{v_0 + v} E_x^i - \frac{ik_x(\epsilon - 1)(v_1 - v)}{(v_0 + v)v_1} \frac{2v_0 \, E_z^i}{\epsilon v_0 + v + (\epsilon - 1)k^2/v_1} \tag{51}$$

We find again that the corrections in $\operatorname{Im} \alpha_{xx}$ to the local theory (see Eq.(A17)) are smaller than $D_0$ by factors $O(a_0/z_0, z_0/\lambda)$, hence they are negligible.

We conclude that this "ideal" surface differs from the surface charge layer of the previous subsection. The charge layer model averages on the short scale $a_0$ and represents in some sense a rough surface, at which the surface charge is found in a self-consistent way. We have checked that the reflection coefficients found here are in agreement with the approach of Kliewer and Fuchs (see, e.g., Refs.9, 10), provided one uses the nonlocal dielectric function for the bulk mentioned after Eq.(8). A very similar calculation with the same results recently appeared in Ref.12.

## VI. DISCUSSION

Charge diffusion in the surface layer is seen to increase the fluctuations of electric fields. In particular at large distances, it changes the distance dependence of the noise power from $1/z_0^2$ into $1/z_0^4$ or $1/z_0^6$, depending on the field polarization. This result may account for the anomalous heating observed in cold ion experiments [5, 6, 7, 8]. In particular Refs.6 and 5 find power laws $1/z_0^n$ for the electric noise with exponents between $n = 3.47 \pm 0.16$ and $n = 3.8 \pm 0.6$, consistent with our prediction $1/z_0^4$ from Eq.(41). The heating rate is $10^2 - 10^3$ higher than what is expected from thermal noise. This would imply enhanced surface diffusion (dimensionless coefficient $D_0 \gg 1$). These experiments also find a frequency dependence $S(\omega) \sim \omega^{-0.8 \pm 0.4}$ and $S(\omega) \sim 1/\omega$, respectively, which is marginally consistent with $\omega^{-1/2}$ of Eq.(41) if the correction in $D_0$ dominates [note the Bose factor $\sim T/\omega$ from Eq.(2)]. More recent data[8] claim, however, a spectrum $S(\omega) \sim \omega^{-1.4 \pm 0.4}$ at fixed $z_0 = 40\mu$m and a noise level smaller by about two orders of magnitude compared with Ref.6.

The model of a charge layer represents an average over details of the charge distribution, and in this sense it is a model of a rough surface. It may corresponds to the "patch" model [5, 6] where random metallic segments contribute to the noise. The diffusion constant $D$ represents correlated charge fluctuations on a scale $\sim \sqrt{D/\omega}$ which could be a patch size. To account for the magnitude of the noise, our parameter $D_0$ needs to be large. A study of the surface diffusion is needed to identify the value of $D_0$. An alternative scenario leading to a surface charge layer is the presence of surface electronic bands in certain surfaces of certain metals [13, 14]. These surface states have led to the recent discovery of acoustic surface plasmons [14]. In both cases, the heating rates observed in ion microtraps may be used as a probe of enhanced surface diffusion.

In a separate work we have studied surface plasmons by using similar NLED models [19]. Furthermore, surface plasmons yield a small but finite electromagnetic noise even for superconductors at temperatures well below their critical temperature.

## APPENDIX A: LOCAL ELECTRODYNAMIC THEORY

For completeness, we rederive here the results for a local $\epsilon(\omega)$, i.e. $D = D_s = 0$.

Within the metal or the vacuum $\epsilon(\omega)$ is uniform so that $\boldsymbol{\nabla} \cdot \mathbf{E} = 0$, however at the boundary we have from Eq.(18) (with $D = 0$) $\boldsymbol{\nabla}[\epsilon(z)\mathbf{E}] = 0$, where $\epsilon(z)$ jumps from $\epsilon$ to 1. Integrating from $z = -0$ to $z = +0$ gives $E_z^{out} = \epsilon E_z^{in}$. As usual, $\oint \mathbf{E} \cdot d\ell = 0$ yields $E_x^{out} = E_x^{in}$, so that

$$\int_{-0}^{+0} \boldsymbol{\nabla} \cdot \mathbf{E} = E_z^{out} - E_z^{in} = (\epsilon - 1)E_z^{in}. \tag{A1}$$

Hence there is a charge sheet at z=0 with

$$\boldsymbol{\nabla} \cdot \mathbf{E} = (\epsilon - 1)E_z^{in}(x,y)\delta(z) \tag{A2}$$

From Eq.(18), we have

$$\nabla^2 \mathbf{E} + \frac{\omega^2}{c^2}\epsilon \mathbf{E} = \boldsymbol{\nabla}[\boldsymbol{\nabla} \cdot \mathbf{E}] = (\epsilon - 1)\boldsymbol{\nabla}[\delta(z)E_z^{in}]. \tag{A3}$$

Integrating the $z$ component across the boundary yields $\partial_z E_z^{out} - \partial_z E_z^{in} = 0$, while integrating the $x$ component yields a jump in $\partial_z E_x$, summarized in the following boundary conditions:

$$\begin{aligned} E_{x,y}^{in} &= E_{x,y}^{out}, \qquad \epsilon E_z^{in} = E_z^{out} \\ \partial_z E_{x,y}^{out} - \partial_z E_{x,y}^{in} &= (\epsilon - 1)\partial_{x,y}E_z^{in}, \qquad \partial_z E_z^{in} = \partial_z E_z^{out}. \end{aligned} \tag{A4}$$

The jump in $\partial_z E_x$ is actually equivalent to the continuity of the magnetic field.

Proceeding from Eq.(27) we consider first the $z$ component with boundary conditions

$$\begin{aligned} \epsilon E_z^t(\mathbf{k}) &= E_z^i(\mathbf{k}) + E_z^r(\mathbf{k}) \\ v E_z^t(\mathbf{k}) &= v_0[E_z^i(\mathbf{k}) - E_z^r(\mathbf{k})] \end{aligned} \tag{A5}$$

The solution for the reflected wave contains the Fresnel coefficient in TM-polarization,

$$E_z^r(\mathbf{k}) = \frac{\epsilon v_0 - v}{\epsilon v_0 + v} E_z^i(\mathbf{k}) \tag{A6}$$

so that the reflected field (6) becomes

$$E_z^r(\mathbf{r}_0) = -2\pi \int \frac{d^2k}{(2\pi)^2} \frac{\epsilon v_0 - v}{\epsilon v_0 + v}[-i(\mathbf{k}_i \cdot \mathbf{a})v_0 - \frac{\omega^2}{c^2}a_z]\frac{e^{-2v_0 z_0}}{v_0}. \tag{A7}$$

Terms odd in $k_x, k_y$ vanish, hence the final form is

$$E_z^r(\mathbf{r}_0) = a_z \int_0^\infty dk\, k^3 \frac{\epsilon v_0 - v}{\epsilon v_0 + v}\frac{e^{-2v_0 z_0}}{v_0}. \tag{A8}$$

As a concrete application we use the form

$$\epsilon(\omega) = 1 + 4\pi\sigma i/\omega \tag{A9}$$

and define the skin depth $\delta = c/\sqrt{2\pi\sigma\omega}$, so that for low frequencies $\omega \ll \sigma$

$$v = \sqrt{k^2 - \frac{\omega^2}{c^2}\epsilon(\omega)} \approx \sqrt{k^2 - \frac{2i}{\delta^2}} \tag{A10}$$

The contribution of $k \lesssim \omega/c$ is $\sim e^{-bz_0/\lambda}$ where $b = O(1)$ and the radiation wavelength is $\lambda = 2\pi c/\omega$, showing an exponential decay at $z_0 \gg \lambda$. In the following we assume $z_0 \ll \lambda$ and consider two regimes:

(i) $\delta \ll z_0 \ll \lambda$: The dominant integration range has $k \approx 1/z_0 \ll \epsilon\omega/c$ and $v \approx \frac{1-i}{\delta}$ (using $\text{Re}(v) > 0$). Expansion in $1/\epsilon$ yields

$$E_z^{r(1)} = a_z \int_0^\infty dk\, k^2 [1 - \frac{v}{k\epsilon}] e^{-2kz_0} = a_z [\frac{1}{4z_0^3} + \frac{(1+i)\delta\omega^2}{4z_0^2 c^2}] \tag{A11}$$

(ii) $z_0 \ll \delta \ll \lambda$: Here $k \approx 1/z_0 \gg \epsilon\omega/c$ so that

$$E_z^{r(2)} = a_z \int_0^\infty dk\, k^2 \frac{\epsilon-1}{\epsilon+1} e^{-2kz_0} = a_z(1 - \frac{1}{\epsilon z_0^2}) \tag{A12}$$

Hence

$$\begin{aligned} \text{Im}\, \alpha_{zz} &= \frac{\omega}{8\pi\sigma}\frac{1}{\delta z_0^2} & \delta \ll z_0 \\ &= \frac{\omega}{8\pi\sigma}\frac{1}{z_0^3} & z_0 \ll \delta\,. \end{aligned} \tag{A13}$$

Consider next the $x$ component with boundary conditions (A4)

$$\begin{aligned} E_x^i + E_x^r &= E_x^t \\ v_0(E_x^i - E_x^r) - vE_x^t &= (\epsilon - 1)ik_x E_z^t = (1 - \frac{1}{\epsilon})ik_x \frac{2\epsilon v_0}{\epsilon v_0 + v} E_z^i \end{aligned} \tag{A14}$$

Eliminating $E_x^r$ yields

$$E_x^r(\mathbf{r}_0) = a_x \int_0^\infty dk\, k\frac{e^{-2v_0 z_0}}{v_0} [\tfrac{1}{2}v_0^2 \frac{v - v_0}{v + v_0} + \frac{k^2 v_0^2(\epsilon - 1)}{(v + v_0)(\epsilon v_0 + v)} - \tfrac{1}{2}\frac{\omega^2}{c^2}\frac{v - v_0}{v + v_0}] \tag{A15}$$

This form is rather tricky for expansion, due to cancellations between the first two terms. We therefore rewrite these as

$$\frac{v - v_0}{v + v_0} + \frac{2k^2(\epsilon - 1)}{(v + v_0)(\epsilon v_0 + v)} = \frac{v^2 + \epsilon v_0 v - v_0 v - \epsilon v_0^2 + 2k^2(\epsilon - 1)}{(v + v_0)(\epsilon v_0 + v)} = \frac{\epsilon v_0 - v}{\epsilon v_0 + v} \tag{A16}$$

using $2k^2(\epsilon - 1) = 2\epsilon v_0^2 - 2v^2$. Hence

$$E_x^r(\mathbf{r}_0) = \tfrac{1}{2}a_x \int_0^\infty dk\, k\frac{e^{-2v_0 z_0}}{v_0}[v_0^2 \frac{\epsilon v_0 - v}{\epsilon v_0 + v} - \frac{\omega^2}{c^2}\frac{v - v_0}{v + v_0}] \tag{A17}$$

as in Ref.5 (see their Eq.(A5)). Expansion as above yields

$$\begin{aligned}\text{Im}\,\alpha_{xx} &= \frac{\omega}{8\pi\sigma}\frac{1}{\delta z_0^2} & \delta \ll z_0 \\ &= \frac{\omega}{16\pi\sigma}\frac{1}{z_0^3} & z_0 \ll \delta\,.\end{aligned} \quad (A18)$$

The form for $\delta \ll z_0$ differs from Ref.5 [Eq.(A7)] by a factor 2, but it does agree with Ref.4 [Eq.(20)].

## APPENDIX B: MAGNETIC FLUCTUATIONS

Magnetic fluctuations from a metallic surface are well studied [2, 3] and the data is in good agreement with the local theory [4, 16]. It is therefore important to study the effects of NLED on the magnetic fluctuations and check if the electric noise enhancement when $D_s/D \gg 1$ is still consistent with a negligible magnetic noise due to the diffusion $D_s$.

Consider a magnetic moment $\mathbf{m}\delta^3(\mathbf{r}-\mathbf{r}_0)$ as a source of radiation with frequency $\omega$. This source is equivalent to a current source $J = \boldsymbol{\nabla} \times [\mathbf{m}\delta^3(\mathbf{r}-\mathbf{r}_0)]$, which from Eq.(4) yields an incoming electric field at $0 < z < z_0$,

$$\mathbf{E}_i(\mathbf{k}) = -\frac{2\pi\omega}{v_0 c^2}\mathbf{k}_i \times \mathbf{m}\,\mathrm{e}^{-i\mathbf{k}_i\cdot\mathbf{r}_0} \quad (B1)$$

It is convenient to work with the boundary conditions of electric fields as in section IV, and at the end find the reflected magnetic field $\mathbf{B}^r(\mathbf{r}_0)$. The boundary conditions for $E_z$, Eq.(22), yields

$$E_z^r(\mathbf{k}) = -\frac{2\pi\omega}{v_0 c^2}(k_x m_y - k_y m_x)\frac{\tilde{\epsilon}v_0 - v}{\tilde{\epsilon}v_0 + v}\mathrm{e}^{-i\mathbf{k}_i\cdot\mathbf{r}_0} \quad (B2)$$

while for the $x$ component Eq.(35) yields

$$E_x^r(\mathbf{k}) = -\frac{v-v_0}{v+v_0}E_x^i - \frac{ik_x}{v+v_0}\frac{\epsilon-1}{\tilde{\epsilon}v_0+v}\frac{2v_0}{\tilde{\epsilon}v_0+v}E_z^i \quad (B3)$$

where $\omega D/c^2 \ll 1$ is neglected. The reflected field is then

$$B_z^r(\mathbf{k}) = \frac{c}{\omega}(k_x E_y^r - k_y E_x^r) = -k^2\frac{v-v_0}{v+v_0}\frac{2\pi m_z}{v_0 c}\mathrm{e}^{-i\mathbf{k}_i\cdot\mathbf{r}_0} + \text{terms odd in }k_x\text{ or }k_y \quad (B4)$$

The odd terms vanish in the $\mathbf{k}$ integration, hence the magnetic response function is

$$\mathcal{B}_{zz} = \frac{B_z^r(\mathbf{r}_0)}{m_z} = \int_0^\infty dk\,\frac{k^3}{cv_0}\frac{v_0-v}{v_0+v}\mathrm{e}^{-2v_0 z_0} \quad (B5)$$

which is identical with the result for the local theory [4, 16].

For the reflected $x$-component, we have

$$\begin{aligned}B_x^r(\mathbf{k}) &= \frac{c}{\omega}(k_y E_z^r - iv_0 E_y^r) \\ &= \frac{2\pi m_x}{cv_0}\left[\frac{k_y^2}{\tilde{\epsilon}v_0+v}\left(\tilde{\epsilon}v_0 - v - 2v_0^2\frac{\epsilon-1}{1+\frac{iD_s k^2}{\omega}}\frac{1}{v+v_0}\right) - v_0^2\frac{v-v_0}{v+v_0}\right]\mathrm{e}^{-i\mathbf{k}_i\cdot\mathbf{r}_0}\end{aligned} \quad (B6)$$

up to terms odd in $k_y$ that vanish in the **k** integration. The change in the response function $\Delta \mathcal{B}_{xx} = \mathcal{B}_{xx} - \mathcal{B}_{xx}(D_s = 0)$ is

$$\operatorname{Im} \Delta \mathcal{B}_{xx} = \operatorname{Im} \int_0^\infty \frac{k^3}{cv_0} \frac{iD_s k^2/\omega}{v + v_0} \frac{-vv_0(\epsilon v + v_0)}{(\epsilon v_0 + v(1 + iD_s k^2/\omega))(\epsilon v_0 + v)} e^{-2v_0 z_0} . \tag{B7}$$

For $\delta \ll z_0 \ll \lambda$ we have $v_0 \approx k$, $v \approx \frac{1-i}{\delta} \gg k \approx 1/z_0$, $\epsilon v_0 \gg v$, hence

$$\operatorname{Im} \Delta \mathcal{B}_{xx} \approx \frac{D_s}{c\sigma} \frac{1}{\delta z_0^4} \tag{B8}$$

which is smaller then the local theory result [4, 16] $\delta/c z_0^4$ by a factor of $a_0^2 D_s/\delta^2 D \approx 10^{-12} D_s/D$ with a typical $\delta \approx 100\mu$. For $\delta \gg z_0$ we have $v \approx k(1 - i/\delta^2 k^2)$, hence

$$\operatorname{Im} \Delta \mathcal{B}_{xx} \approx \frac{D_s}{c\sigma} \frac{1}{\delta^2 z_0^3} \tag{B9}$$

which is small compared with the local theory result $1/cz_0\delta^2$ by a factor of $D_0\omega\delta^2/\sigma z_0^2 \approx 10^{-12} D_0$ with $\delta \sim z_0$. We conclude that even if $D_0$ is sufficiently large to account for the observed electric noise, e.g. $D_0 \sim 10^3 - 10^4$, the effect on the magnetic noise is negligible and therefore our NLED model of a charge layer is consistent with the magnetic noise data [2, 3].

**Acknowledgment:**

We thank Ron Folman for discussions and helpful communications. One of us (BH) thanks E. V. Chulkov and V. M. Silkin for illuminating discussions. This research was supported by THE ISRAEL SCIENCE FOUNDATION founded by the Israel Academy of Sciences and Humanities, by the Deutsch-Israelische Projektkooperation (DIP), and by the European Union (project QUELE within the IST programme, contract no. 003772).